%% file: main.tex
\documentclass{hotnets17}

\usepackage{times}
\usepackage[english]{babel}
\usepackage{booktabs}
\usepackage{paralist}
\usepackage{amsmath}
\usepackage{graphicx}
\usepackage{hyperref}
\usepackage{url}
\usepackage{epstopdf}
\usepackage{cite}
\usepackage{enumerate}
\usepackage{fixltx2e}
\hypersetup{pdfstartview=FitH,pdfpagelayout=SinglePage}

\usepackage[labelformat=simple, subrefformat=simple]{subcaption}

\title{More Than The Sum Of Its Parts: Exploiting\\ Cross-Layer and Joint-Flow Information in MPTCP}

\begin{document}

\numberofauthors{1}
\author{
  \alignauthor Tanya Shreedhar$^\dag$, Nitinder Mohan$^\ddag$, Sanjit K. Kaul$^\dag$ and Jussi Kangasharju$^\ddag$\\
    \affaddr{$^\dag$IIIT-Delhi, India \hspace{2em} $^\ddag$University of Helsinki, Finland}    
%  \alignauthor Nitinder Mohan\\
%    \affaddr{University of Helsinki, Finland}  
%    \and
%  \alignauthor Sanjit K. Kaul\\
%    \affaddr{IIIT-Delhi, India}
%  \alignauthor Jussi Kangasharju\\
%     \affaddr{University of Helsinki, Finland}
}

\maketitle

\begin{abstract}
Multipath TCP (MPTCP) is an extension to TCP which aggregates multiple parallel connections over available network interfaces. MPTCP bases its scheduling decisions on the individual RTT values observed at the subflows, but does not attempt to perform any kind of joint optimization over the subflows. Using the MPTCP scheduler as an example, in this paper we demonstrate that exploiting cross-layer information and optimizing scheduling decisions jointly over the multiple flows, can lead to significant performance gains. While our results only represent a single data point, they illustrate the need to look at MPTCP from a more holistic point of view and not treat the connections separately, as is currently being done. We call for new approaches and research into how multiple parallel connections offered by MPTCP should be used in an efficient and fair manner.

%  To efficiently schedule application data over TCP subflows, MPTCP relies on smoothed RTT (SRTT) observed at each subflow. In this paper, we show that the current MPTCP implementation and research consider each available subflow independently and therefore provides sub-optimal performance over available capacity. Through our real world experimentation, we demonstrate that MPTCP does not take a holistic view of the network due to its dependence on TCP-level information.  To prove our claim, we consider current occupancy of device queue in MPTCP scheduling decisions. We evaluate the correctness of our model over MATLAB and implement it in realistic scenarios involving WiFi and cellular in NS3. We argue that MPTCP must consider certain cross-layer/cross-flow information and conditions, also presented in the paper, to handle associated subflows efficiently.
\end{abstract}

%\begin{ACMkeywords}
%MultiPath TCP (MPTCP), Scheduler, SRTT, Device queue, Queuing 
%\end{ACMkeywords}

%add all sections here. Make seperate tex files for them
%Section list:
% 1. Introduction
% 2. Related Work
% 3. Device buffer queueing [real-world implementation] 
% 4. Queue-Scheduler (name) [Math+Matlab correctness]
% 5. Evaluation [ns3]
% 6. Conclusion 
% Section End

%%%%%% Vision that we sell %%%%
%% MPTCP has been going in the wrong direction by considering network as N-TCP connections and utilizing the same methodology. It need to consider several other parameters to efficiently utilize all available credible paths.
%% List some parameters that MPTCP needs to consider: at sender, at receiver, within the paths etc. 
%% Explain with diagram queueing in network stack and how TCP deals with it, FIFO queue, CoDEL, Queue Disciplines. Queueing delay is reflected in SRTT of TCP to perform TCP rate control as no other option is available. 
%%Look for papers that are considering MPTCP to work in n-TCP connections (using TCP flows as backup). 

\input{introduction}

\input{device_buffer}
\input{scheduler}

\input{evaluation}

\input{results}

\input{relatedwork}

\input{discussion}

\input{conclusion}

\bibliographystyle{abbrv} 
\begin{small}
\bibliography{hotnets17}
\end{small}

\end{document}

%% file: introduction.tex
\section{Introduction}\label{sec:introduction}

 The Transport Control Protocol (TCP) is an essential part of the Internet and is used by a majority of applications to reliably
transport their data over the network.
 Since its inception, TCP has seen a lot of research, particularly regarding how its performance and throughput could be improved \cite{tcpratemodel, van_red, tcp_throughput}.
 More recent work has led to the development of Multipath TCP (MPTCP), standardized by IETF, which is an extension to standard TCP. It operates under the viewpoint that end hosts like smartphones,  servers, etc., are equipped with multiple access interfaces such as Ethernet, WiFi, 3G/4G, Bluetooth and can form multiple parallel network paths between communicating devices. Unlike a single path TCP connection, MPTCP utilizes multiple parallel TCP subflows between hosts via (all) available network interfaces for
simultaneous data transfer. It achieves robustness and resilience to link failures, provides
seamless connection handovers over different network interfaces and
aggregates throughput and bandwidth of the underlying TCP connections~\cite{mptcpoverview,mptcphard}.
 Due to the benefits offered by MPTCP, researchers have proposed
utilizing it in datacenters~\cite{datacentermptcp}, opportunistic networks~\cite{opportunisticmptcp} etc.
 The Linux implementation of the protocol has been made open-source to
network researchers~\cite{mptcplinux} and a commercial version is being
used by Apple Inc. to support its digital assistant Siri in iOS and MacOS
systems. Recently, Apple Inc. has also made its MPTCP API open to iOS application developers such that they can fully utilize its capability~\cite{applemptcp}.
 
\begin{figure}[!t]             
\centering
\captionsetup{justification=centering}
\includegraphics[width=0.4\textwidth]{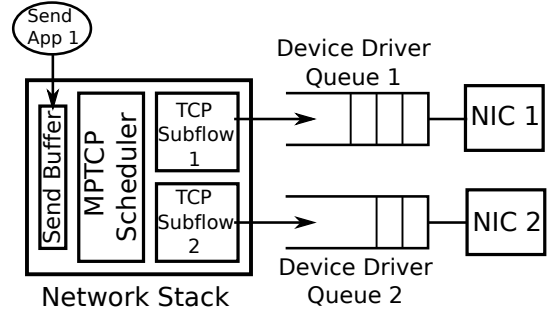}
\caption{\label{fig:mptcp_device_queue}Queues in MPTCP transmission path}
\end{figure}

%What MPTCP does and why it is important 

MPTCP protocol design is
influenced by network compatibility and ensuring fairness to existing
TCP connections.
 Figure~\ref{fig:mptcp_device_queue} shows the network stack of an
MPTCP-compliant client.
 MPTCP adds a scheduling layer over existing TCP connections and
routes application packets to one of the subflows based on a decision
parameter.
 Efficient scheduling decisions can improve the delay performance of MPTCP.
 Several schedulers have been proposed by researchers but due to its
compliance with modular TCP design, the default scheduler injects
packets on a subflow with the lowest smoothed TCP RTT (SRTT) value~\cite{mptcpscheduler}.
 TCP RTT presents delayed information of the internal network behavior
such as congestion, packet drops, queueing, etc.
 and does not explicitly indicate the reason for the change in network
conditions.
 Unlike traditional TCP, MPTCP has the capability to proactively
switch TCP flows if it senses any issues on one of the flows.
However, MPTCP primarily treats the individual streams as separate entities and does not attempt to optimize performance across them.

%  We argue that MPTCP must also consider several other parameters such
% as from other layers of the stack, in-network explicit notifications, etc.
%  to make more holistic scheduling decisions.

In this paper, we argue that MPTCP should take a more comprehensive view over individual subflows and attempt to optimize the overall performance, as opposed to treating them as separate parallel entities. MPTCP currently is limited to the TCP-level information provided by the individual TCP subflows, but we claim that a more holistic approach is needed to exploit the possibilities offered by multiple flows to the full extent. This implies the need to take a different approach in MPTCP research and in general on how to exploit parallel resources in networking, which can also serve other areas of networking~\cite{ciscowhite, edgesurvey,contentcentricnetwork,datacenternetwork}.

%%usage of MPTCP in real environment
%The scheduler is an important block of Multipath TCP architecture which decides the path the data packet would take. Efficient scheduling decision can improve the delay performance in case of MPTCP. 
%However, the default scheduler available for the real world experiments does not utilise the network resources efficiently. Our preliminary experiments show that the default scheduler does not utilise the sub-flows equally even when the channel conditions, RTTs are similar.  We will discuss these in detail in the subsequent sections. The scheduler function is categorised into two broad categories. First, is the selection of sub-flow from the set of sub-flows with available space in congestion window. Second, how much data is to be scheduled on the selected sub-flow. The goal is to minimise the total delay that the packet sees in the path. 

%Problem in MPTCP research thinking, set in single path

To illustrate one aspect of the problem and to provide motivation for
future work, we present several issues in current
implementation and working of MPTCP due to its reliance on TCP layer
information.
 We show via controlled experiments that the default minSRTT
scheduler of MPTCP essentially forces the protocol to use only one of
the many available flows and thus leads to lower performance.
 To demonstrate the possible performance that can be achieved by
MPTCP, we develop the QueueAware scheduler which considers network
interface device queue size while making scheduling decisions.
 We evaluate and compare the performance of QueueAware with minSRTT
scheduler in several realistic scenarios over WiFi and 4G network
interfaces.
We also sketch how other parameters and network conditions currently ignored by MPTCP could be used for improved efficiency of network communication.

%  We finally propose several other parameters and conditions that MPTCP
% must consider explicitly to provide an overall efficient
% connectivity.

%The major contributions of this paper are as follows:
%\begin{enumerate}
%
%\item In this paper, we use a cross-layer approach to minimise the delay in the network. The MPTCP scheduler sits in the transport layer and is oblivious to the functioning/delays in the lower layers. This mechanism is sufficient for the operation of SPTCP(Single Path TCP). However, due to the availability of multiple usable paths, there is a need for understanding the delay characteristics of the paths. This understanding will help in better utilisation of the available network resources.
%\item We demonstrate the problem with the default scheduler using real world experiments.
%\item We model the problem on the existing queueing theory models.
%\item We validate our results in a simplistic MATLAB simulator that gives promising results. 
%\item We simulate various schedulers in NS3 and compare our results with the default scheduler used in Linux kernel implementation.
%\end{enumerate}
%what we present in the paper
The remainder of the paper is organised as follows. In Section~\ref{sec:incognizance}, we show the gap in performance of the default MPTCP scheduler via controlled experiments. In Section~\ref{sec:scheduler}, we describe the scheduling policy used by QueueAware. Section~\ref{sec:evaluation} describes how we evaluate the efficacy of QueueAware using ns-3 simulations. Section~\ref{sec:results} quantifies gains in performance achieved by QueueAware over the default MPTCP scheduler. We discuss related works in Section~\ref{sec:relatedwork}. In Section~\ref{sec:discussion} we discuss various network parameters and network conditions that are currently overlooked by MPTCP when making scheduling decisions. These motivate future avenues for research. We conclude in Section~\ref{sec:conclusion}.

%%% Local Variables:
%%% mode: latex
%%% TeX-master: "main"
%%% End:

%% file: device_buffer.tex
%\section{MPTCP real-world performance analysis [SKK: Plan to edit]}
\section{Downsides of Ignoring Local Information} \label{sec:incognizance}
\begin{figure}[!t]
\centering
\begin{subfigure}{.23\textwidth}
\centering
  \includegraphics[width=\linewidth]{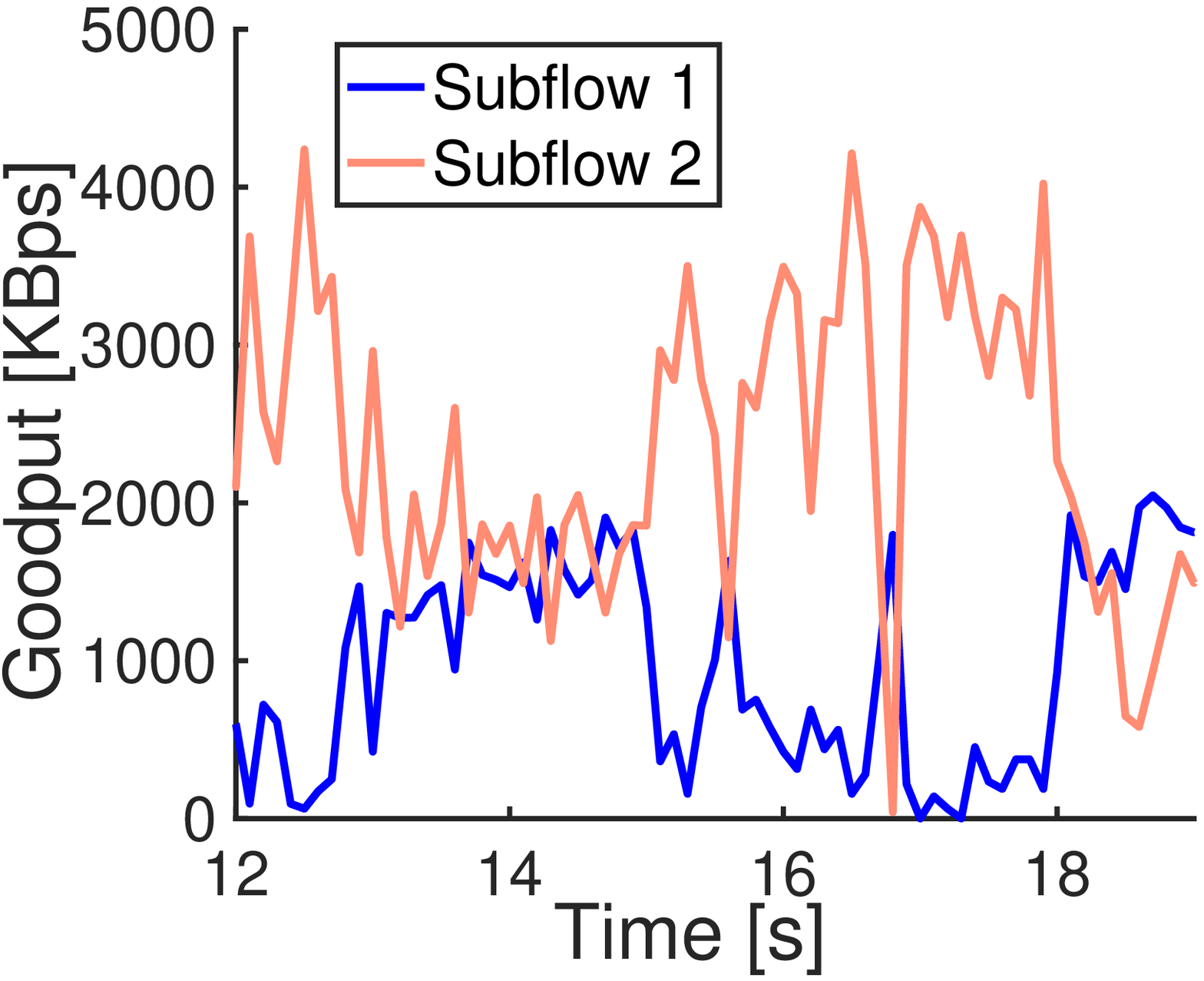} %[width=1.6in]
  \caption{\label{fig:emc_throughput}}
\end{subfigure}%
\hspace{0.005\textwidth}
\begin{subfigure}{.22\textwidth}
\centering
  \includegraphics[width=\linewidth]{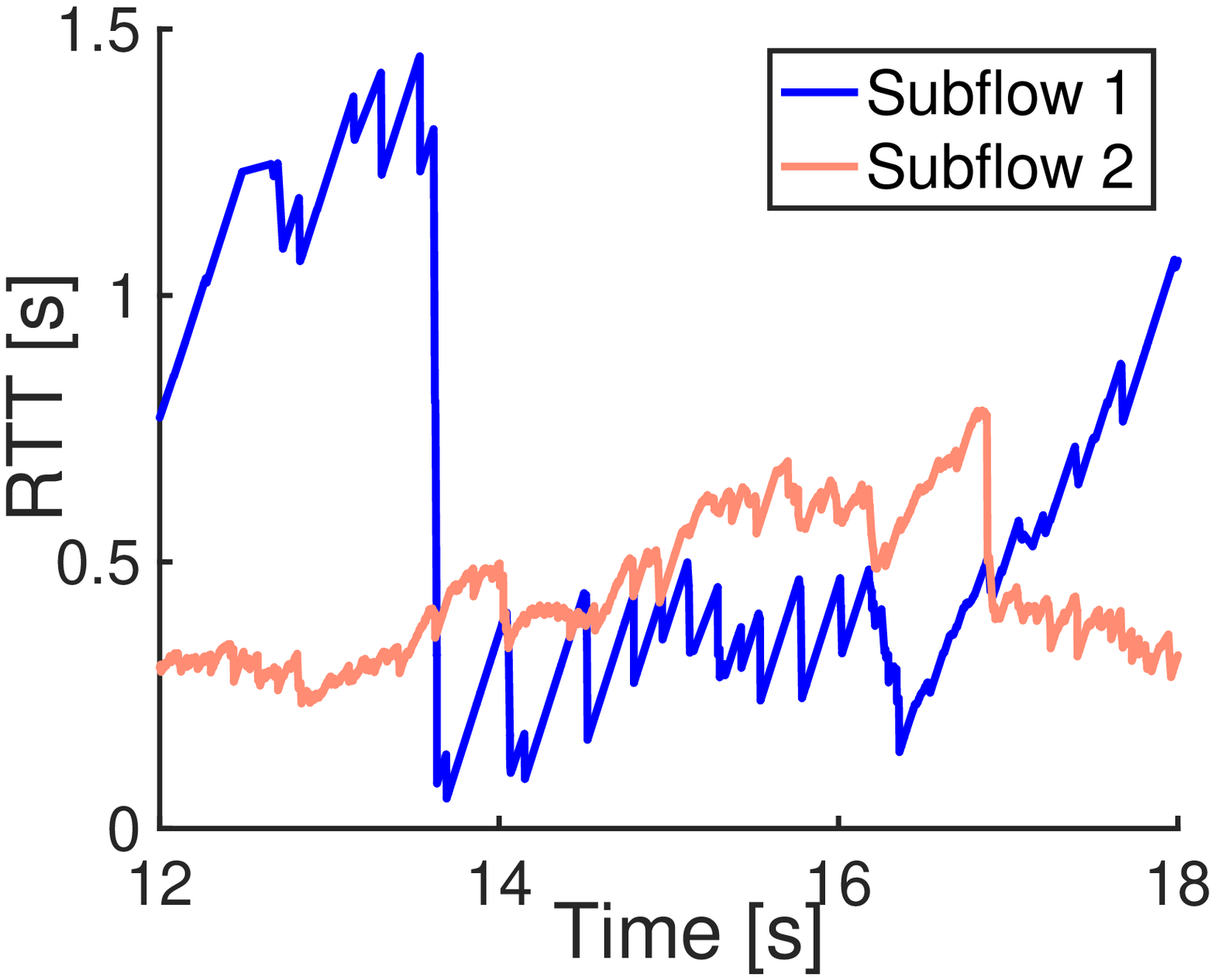} %[width=1.6in]
  \caption{\label{fig:emc_rtt}}
\end{subfigure}
\caption{Experimentally obtained goodputs and RTT(s) of two MPTCP subflows using non-interfering WiFi access paths. The paths and network are shown in Figure~\ref{fig:netw_topology}}
\label{fig:emc_results}
%\vskip -3mm
\end{figure}

\begin{figure}[!t]             
\centering
%\captionsetup{justification=centering}
\includegraphics[width=\columnwidth]{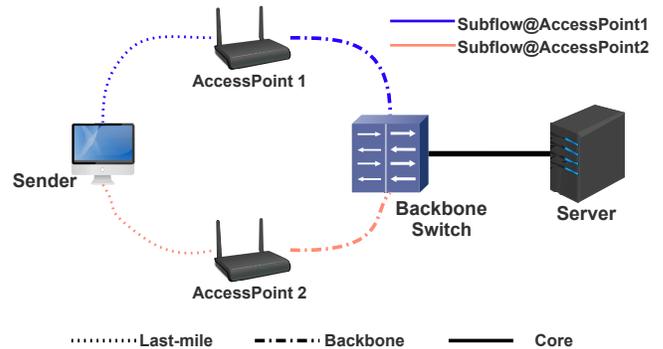}
\caption{Topology used in experiments and simulations}
\label{fig:netw_topology} 
\end{figure}

Figure~\ref{fig:emc_throughput} exemplifies goodput obtained from controlled testbed experiments that show how the default MPTCP scheduler optimizes over two available TCP subflows that use non-interfering end-to-end paths. The network topology used in the experiment is shown in Figure~\ref{fig:netw_topology}. The topology emulates a situation where the access network is a bottleneck, but the core network has high-speed pipes. 

Neither flow drops any packets during the length of the experiment. One would, therefore, expect to fill packets on both flows and achieve stable goodput on both. However, MPTCP more often than not prefers to send packets on one flow over the other and is unable to optimize jointly over the available paths. As a result, MPTCP is only able to utilize $\approx 60\%$ of available aggregated bandwidth in the experiment. We will argue that the reasons for this are two-fold: \emph{(a)} decision making by the scheduler only uses the SRTT values of the flows, which is an end-to-end feedback mechanism and hence delayed; and \emph{(b)} this delayed SRTT feedback leads to the scheduler selecting one flow over the other for undesirably long intervals, instead of suitably allocating packets on both flows. These reasons are in fact a consequence of \emph{MPTCP being incognizant} of cross-layer information that is \emph{readily available locally}.

Interestingly, the scheduling decisions that lead to high device queue occupancy and increase in SRTT were made using values of SRTT that corresponded to an earlier interval when the device queue was lightly loaded. So while a device queue (local to the MPTCP sender and used by an MPTCP flow) is full with packets, MPTCP remains oblivious to the same. Instead, it waits to be informed via a delayed end-to-end SRTT based feedback mechanism about the fact that the flow it has been assigning packets to is in fact loaded. In the process, it loses out on many opportunities of scheduling packets to the other better flow, one experiencing lower queue delay. Further, as the device queue is drop-tail in nature, i.e. it starts dropping incoming packets if full, MPTCP has to retransmit locally dropped packets, the information of which it infers after a missing ACK.  Finally, a rapid dip in SRTT is noticed when a flow, which was earlier heavily queued but stayed unused for a while, is assigned new packets. These packets experience smaller waiting times and thus have a small RTT.

Next, we describe the QueueAware scheduler for MPTCP which is motivated by the observation that MPTCP doesn't have to rely on delayed increase in SRTT to identify local queue delay. We show that instead, using the occupancy of the device queues together with SRTT enables MPTCP to use all available flows more efficiently.

%% file: scheduler.tex
\section{QueueAware Scheduler} \label{sec:scheduler}
\begin{figure}[!t]             
\begin{center}
%\captionsetup{justification=centering}
\includegraphics[width=0.49\textwidth]{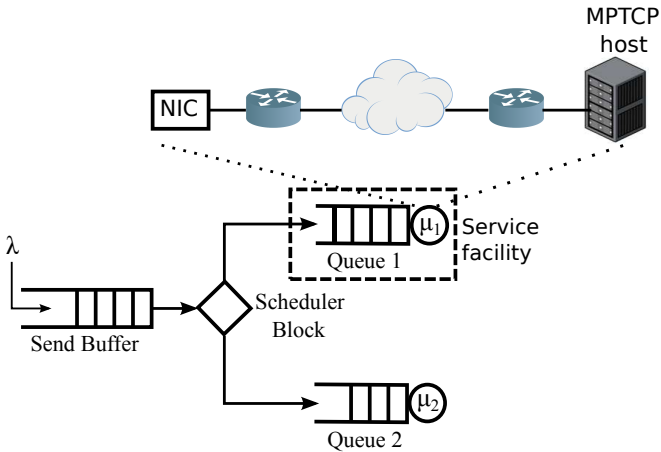}
\caption{Queueing abstraction of an \emph{end-to-end} MPTCP connection with two flows}
\label{fig:networkqueue}
\end{center}
\end{figure}

The MPTCP scheduler chooses amongst one of many available TCP subflows for each application packet arrival, such that the \emph{end-to-end} throughput is maximized. We consider a simplified queue theoretical abstraction to capture the essentials of this problem. Specifically, we model each subflow by a service facility. Figure~\ref{fig:networkqueue} illustrates the abstraction for an MPTCP \emph{end-to-end} connection that uses two TCP flows. The abstraction allows us to apply results from analysis of multi-queue systems\cite{rosberg}.

In our queueing abstraction, packets generated by an application arrive into a queue that models the TCP send buffer (Figure~\ref{fig:mptcp_device_queue}). Packets in this queue are assigned to one of the available service facilities in a first-come-first-serve (FCFS) manner. Each facility consists of a finite queue and a server. Packets inside a facility are serviced in an FCFS manner. 

The queue in a service facility resembles the network interface queue that is used by the TCP subflow corresponding to it. The server includes the network interface card, the destination host (all layers of the TCP/IP stack), and intermittent nodes in the core and the access network used by the flow.

\emph{Origins of the QueueAware scheduler:} Many analytical works on queueing systems have looked at scheduling customers/packets to parallel servers\cite{rosberg,weber,whitt,winston}. Packet arrivals are modeled as a random process (for example, Poisson process) with an arrival rate $\lambda$. The time a server takes to service packets (service time) is modeled as a random variable. Each server has a known service rate, which is the inverse of the expectation of the service time. The two servers in Figure~\ref{fig:networkqueue} have service rates of $\mu_1$ and $\mu_2$. For many general arrival processes and service time distributions, when all servers are stochastically identical, the policy of choosing a service facility with a minimum number of packets, is known to be optimal~\cite{rosberg,weber,winston}, that is it maximizes the number of packets serviced in a given amount of time. For the case of non-identical servers, we were unable to find an optimal policy. However, for the case when the arrivals are Poisson and the service times are non-identical but exponentially distributed, a scheduling policy that assigns a packet to a service facility that minimizes the conditional expected waiting time of the packet, conditioned on the knowledge of the number of packets waiting for service in the facility, is shown to have good performance~\cite{rosberg}. Our QueueAware scheduler uses the policy in an MPTCP setting. 
%First, we state the policy in a precise manner.

Consider $K$ service facilities indexed $1,\ldots,K$. Let facility $k$ have a service rate of $\mu_k$. Let $n_k(t)$ be the number of packets waiting for service in facility $k$ at time $t$. The policy assigns a packet to a service facility $k^*$ given by
\begin{align}
k^* = \arg \min_k \frac{n_k(t)}{\mu_k}.
\label{eqn:policy1}
\end{align}
Note that $1/\mu_k$ is the expected service time of a packet in facility $k$. As a result, the conditional waiting time of a packet that enters such a facility is ${n_k(t)}{\mu_k}$, which is the sum of the expected service times of the $n_k(t)$ packets currently waiting for service in the facility.

\emph{Adapting policy~(\ref{eqn:policy1}) to multiple end-to-end TCP subflows:} The number of packets $n_k(t)$ in service facility $k$ (corresponding to TCP subflow $k$) is the number of packets in the corresponding device queue and can be obtained. However, we must estimate the service rate $\mu_k$.

Consider the $i$\textsuperscript{th} packet arrival. Let $t_i^s$ be the time the packet is assigned to a service facility. Let $t_i^a$ be the time that a TCP ACK acknowledges receipt of the packet. The RTT of the packet can be denoted as $\text{RTT}_i = t_i^a - t_i^s$. Note that the RTT includes the time this packet waits in the queue in its assigned service facility before it starts service and the time it spends in service. Let the wait time be $W_i$\footnote{For simplicity of exposition we ignore the time a TCP ACK may have to wait in a queue before being sent to the TCP layer.}. This time can be calculated locally at the MPTCP sender. The time $X_i$ that the packet spends in service begins when the packet enters the NIC for transmission and ends when a TCP ACK for the packet is received. Given $W_i$ and $\text{RTT}_i$, we have $X_i = \text{RTT}_i - W_i$. The estimate of the service rate is updated on receipt of a TCP ACK. Let $\hat{S}_k$ be the current estimate of the average service time of facility $k$. On receipt of a TCP ACK for the packet, we update
\begin{align}
\hat{S}_k = \alpha \hat{S}_k + (1-\alpha) X_i,
\end{align}
where $0 < \alpha < 1$ applies appropriate weights to the last estimate of the average and the most current service time. We use $\alpha=0.8$ in this work. The corresponding estimate of the service rate is $1/\hat{S}_k$. At time $t$, QueueAware schedules to the TCP subflow $k^*$ that satisfies
\begin{align}
k^* = \arg \min_k {n_k(t)} \hat{S}_k.
\label{eqn:policy3}
\end{align}

%% file: evaluation.tex
\section{Evaluation Methodology}\label{sec:evaluation}
We simulated network topologies of the kind shown in Figure~\ref{fig:netw_topology} using the network simulator ns-3. An MPTCP client uses two TCP subflows to the MPTCP server. The simulator ns-3 has an implementation of MPTCP~\cite{ns3mptcp} that we modified to include QueueAware. We compare the performance of the default minSRTT scheduler of MPTCP with that of QueueAware. 

We simulated scenarios where both the MPTCP flows have reliable paths, and where one of the paths is unreliable and drops TCP packets. %errors on both paths? 
We will show that QueueAware, unlike minSRTT, is able to jointly use available flow capacities to achieve larger per flow (and aggregate) goodputs. Also, it does \emph{at least} as well as the default minSRTT scheduler in scenarios where paths used by TCP subflows are unreliable. 
%resulting in TCP shrinking its congestion window. 

\textbf{Reliable Paths:} We consider two non-interfering paths that see no packet drops but the corresponding access networks (last mile in Figure~\ref{fig:netw_topology}) have different bottleneck rates. Specifically, we simulated the following network scenarios:
\begin{itemize}
\item \emph{Identical WiFi Access Points:} MPTCP client uses its WiFi interfaces to connect to identical WiFi access points providing reliable links. This use case may occur in an enterprise WiFi network, where a client may have more than one \emph{good} access point in its vicinity. We set the WiFi link rate to each access point to $6$ Mbps.
\item \emph{Non-identical WiFi Access Points:} The use case remains the same as above. However, one of the two access points has a faster link of $12$ Mbps, for example, because of greater proximity to the MPTCP client.
\item \emph{Heterogeneous networks of WiFi and 4G:} MPTCP client uses its WiFi interface and $4$G interface to connect to access points of the different technologies. The WiFi link rate is $6$ Mbps and $4$G link rate is $12$ Mbps. Though the link rate of $4$G is large, packets experience larger RTT over $4$G \cite{mptcpwirelessstudy}.
\end{itemize}

\textbf{Unreliable Paths:} We simulated an MPTCP client that uses its two WiFi interfaces to connect to access points. However, the tcp subflow using one of the access points suffers TCP packet errors. We set packet error rate to $10^{-2}$. In a real setting, this could happen because the client is close to losing coverage from the access point or because the access point is heavily loaded.% We also simulate a scenario, for when the client is leaving WiFi coverage, where both the TCP subflows suffer packet error rates of about $10^{-2}$.

Lastly, for a selection of paths and network technologies, we simulated short time MPTCP subflows by performing a small file upload of $10$ MB.

For all simulations, the backbone link was modeled as a Ethernet with rate $30$ Mbps, and the core network was modeled as a $50$ Mbps link. Our choice of links within the access networks makes them the bottleneck. The application at the MPTCP client generated a load of $50$ Mbps. All results were averaged over 10 simulation runs. The access networks, backbone, and the core see no traffic other than that created by our MPTCP client. We defer performance evaluation of QueueAware under more realistic loads and larger numbers of MPTCP clients to future work.

\vspace{2mm}

%% file: results.tex
\section{Results} \label{sec:results}

\begin{figure*}[!tbp]
\centering
\captionsetup{justification=centering}
\begin{subfigure}{.24\textwidth}
  \centering
  \includegraphics[width=\linewidth]{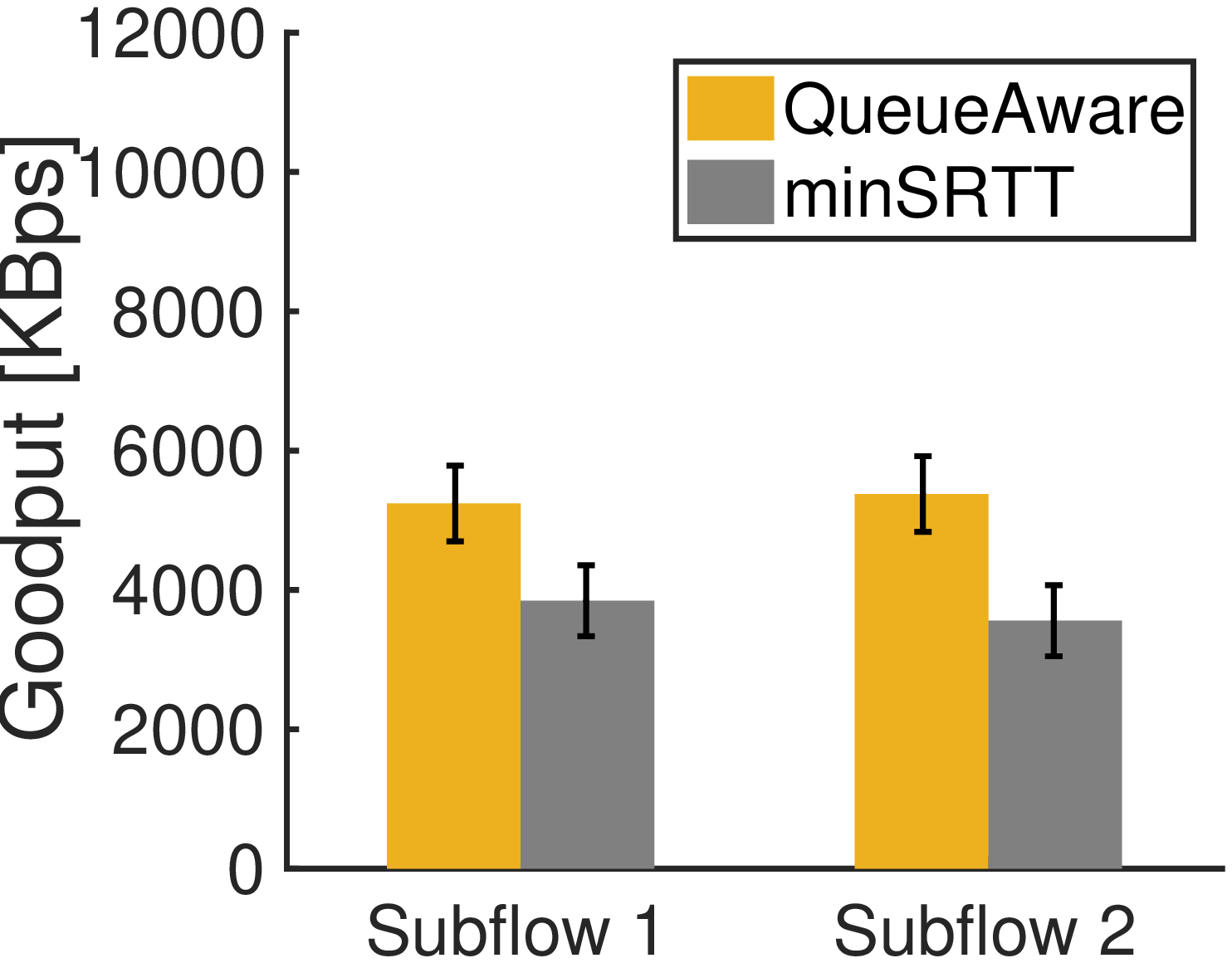}
  \caption{\label{fig:goodput_wlan+wlan} Identical WiFi access points}
\end{subfigure}%
\hspace{0.002\textwidth}
\begin{subfigure}{.24\textwidth}
  \centering
  \includegraphics[width=\linewidth]{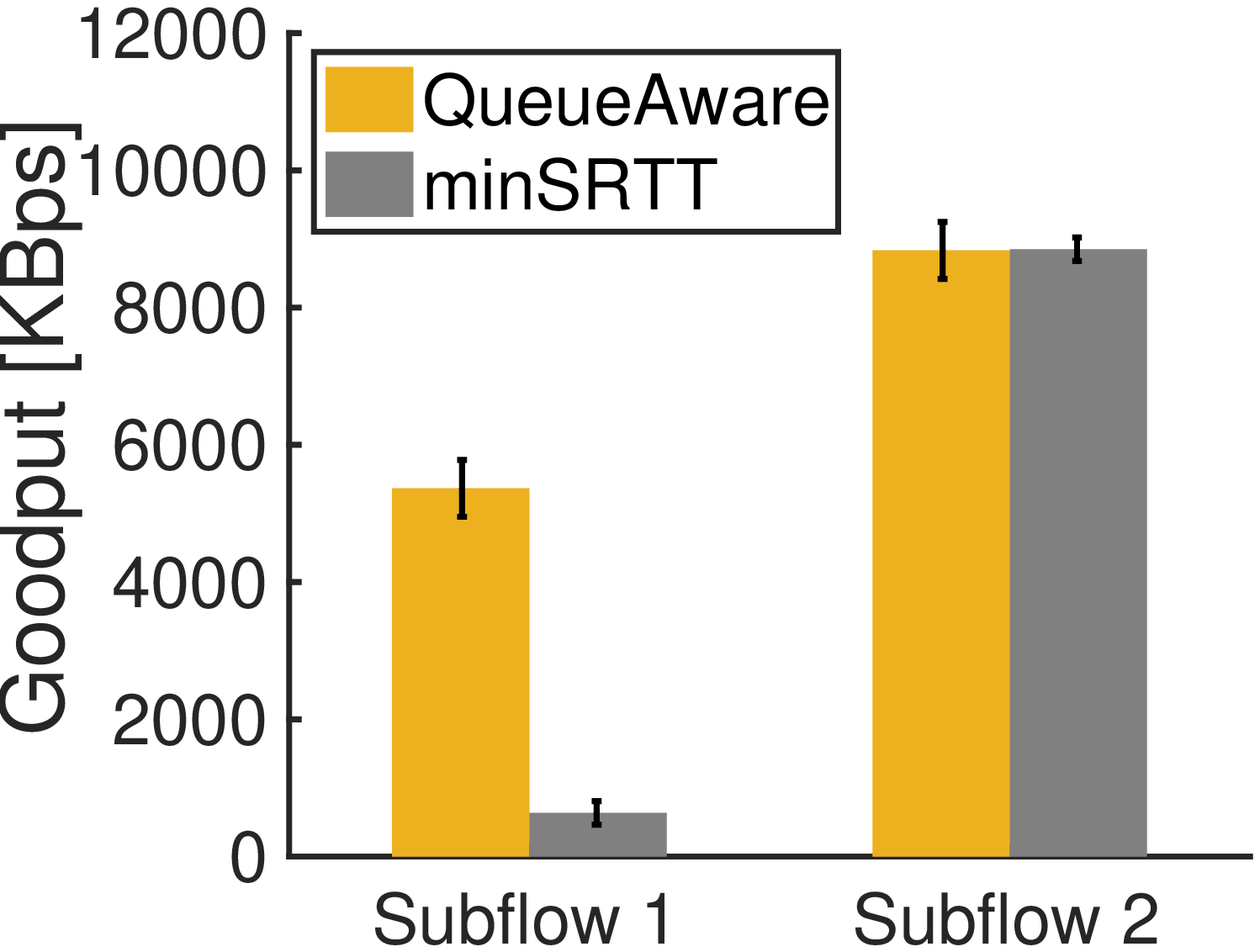}
  \caption{\label{fig:goodput_wlan+wlan_6_12} Non-identical WiFi access points}
\end{subfigure}
\hspace{0.002\textwidth}
\begin{subfigure}{.23\textwidth}
  \centering
  \includegraphics[width=\linewidth]{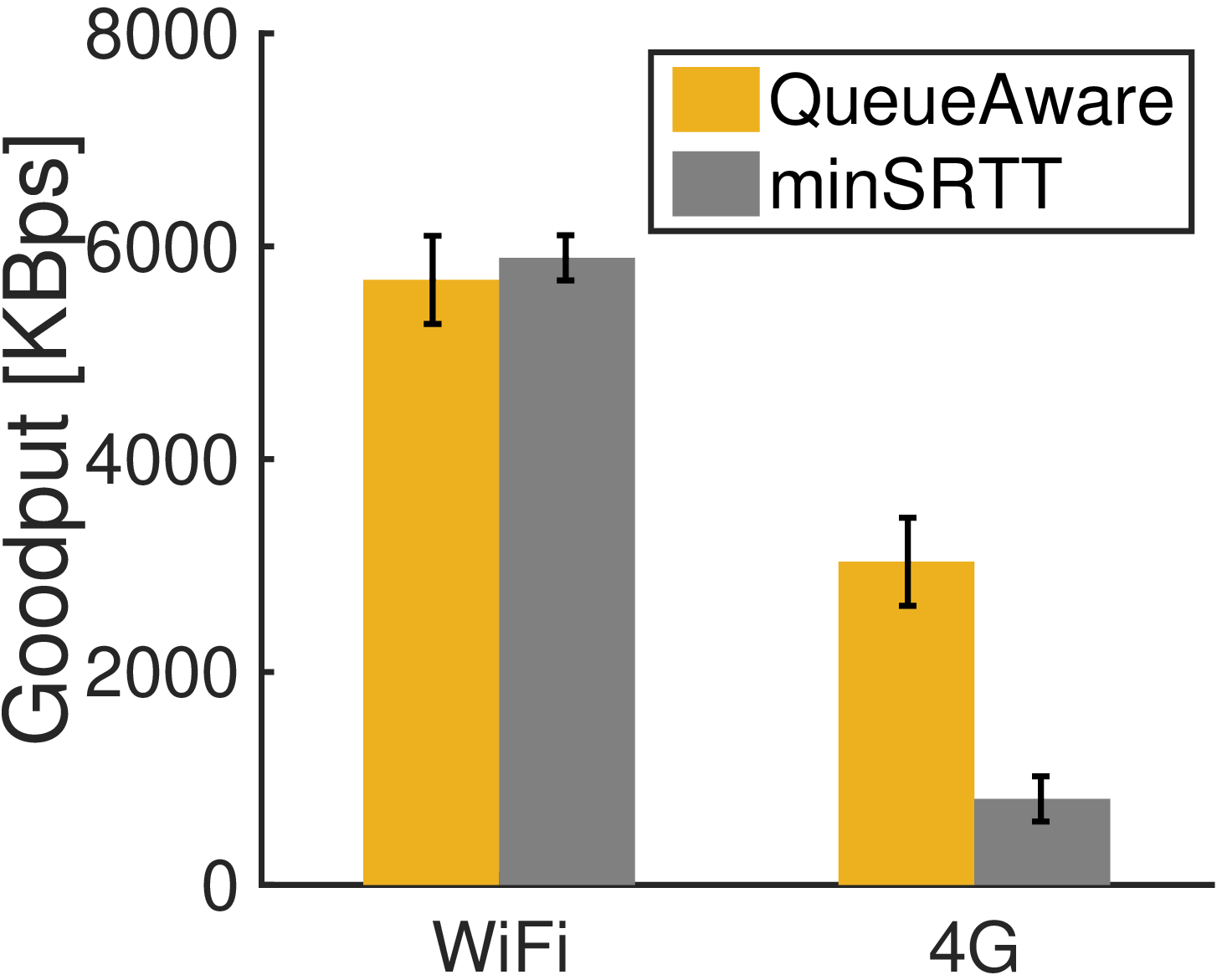}
  \caption{\label{fig:goodput_wlan+lte}WiFi and $4$G.}
\end{subfigure}
\hspace{0.002\textwidth}
\begin{subfigure}{.23\textwidth}
\centering
  \includegraphics[width=1\linewidth]{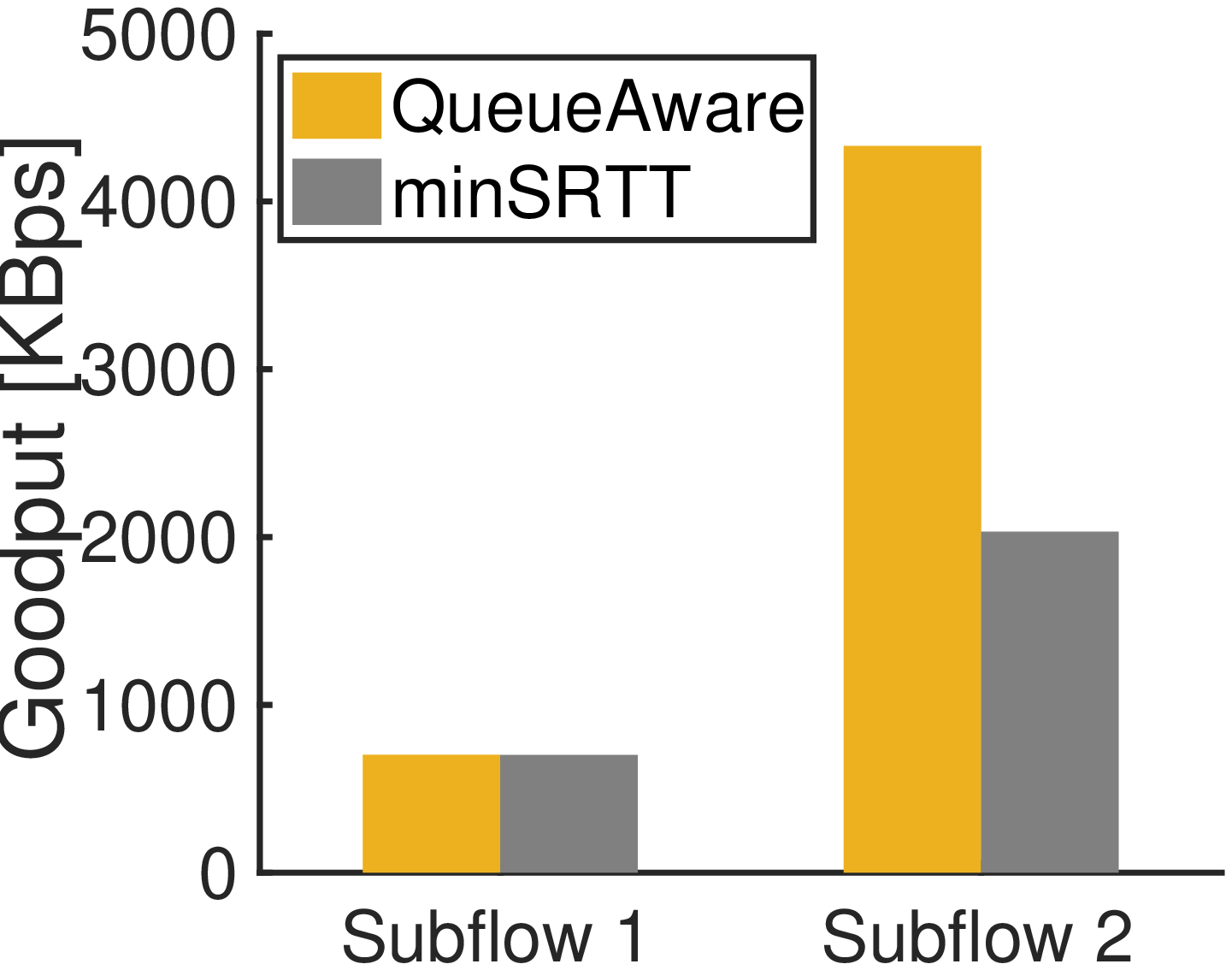}
  \caption{\label{fig:error_2} One unreliable path}
\end{subfigure}
\caption{{Goodputs of MPTCP flows when using QueueAware and minSRTT, for the scenarios described in Section~\ref{sec:evaluation}}}
\label{fig:goodputs_result}
%\vskip -3mm
\end{figure*} 

\subsection{Reliable Paths}
\begin{figure}[!t]
\centering
\begin{subfigure}{.23\textwidth}
\centering
  \includegraphics[width=\linewidth]{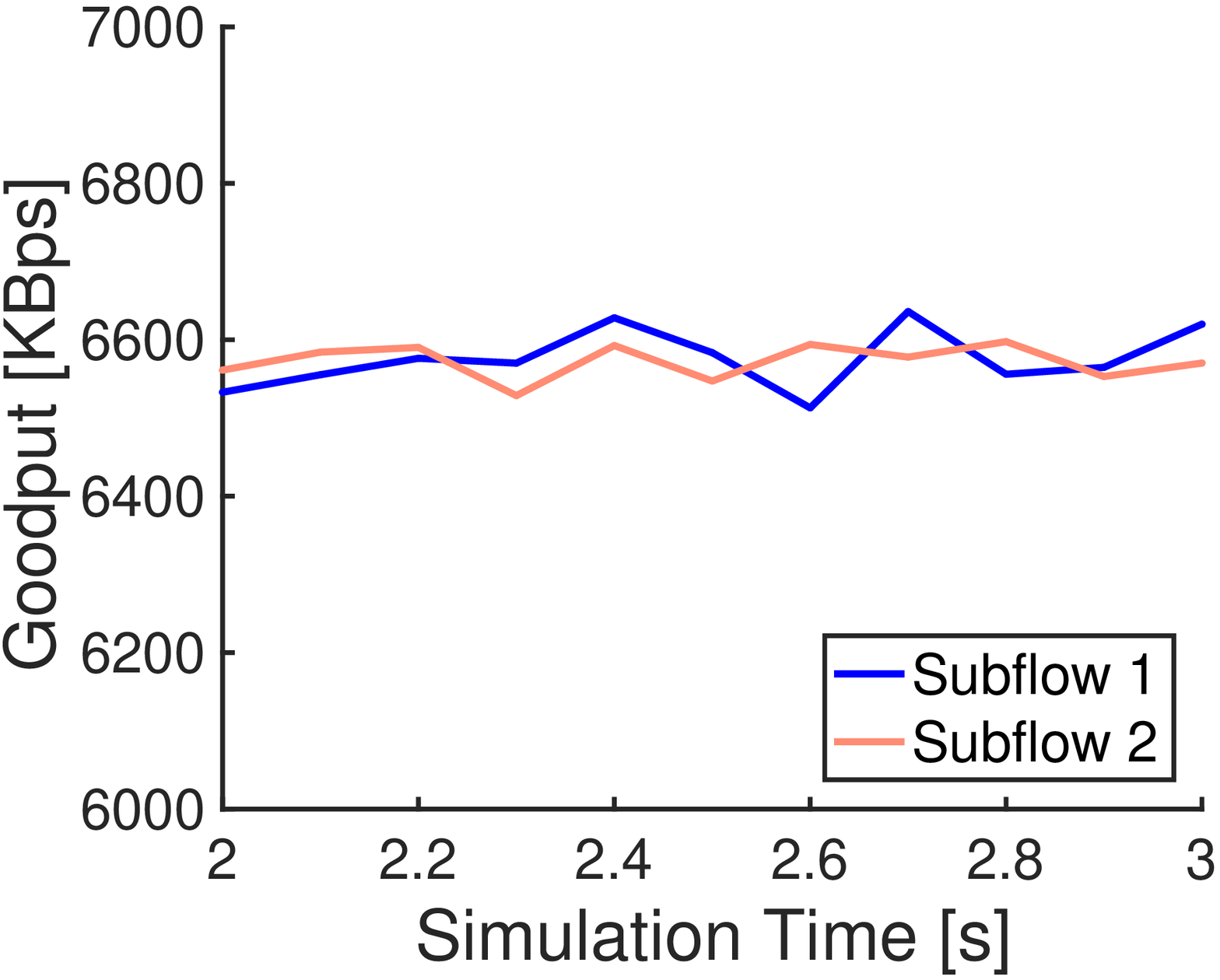}
  \caption{\label{fig:queuesched_throughput_compare}QueueAware}
\end{subfigure}%
\hspace{0.005\textwidth}
\begin{subfigure}{.22\textwidth}
\centering
  \includegraphics[width=\linewidth]{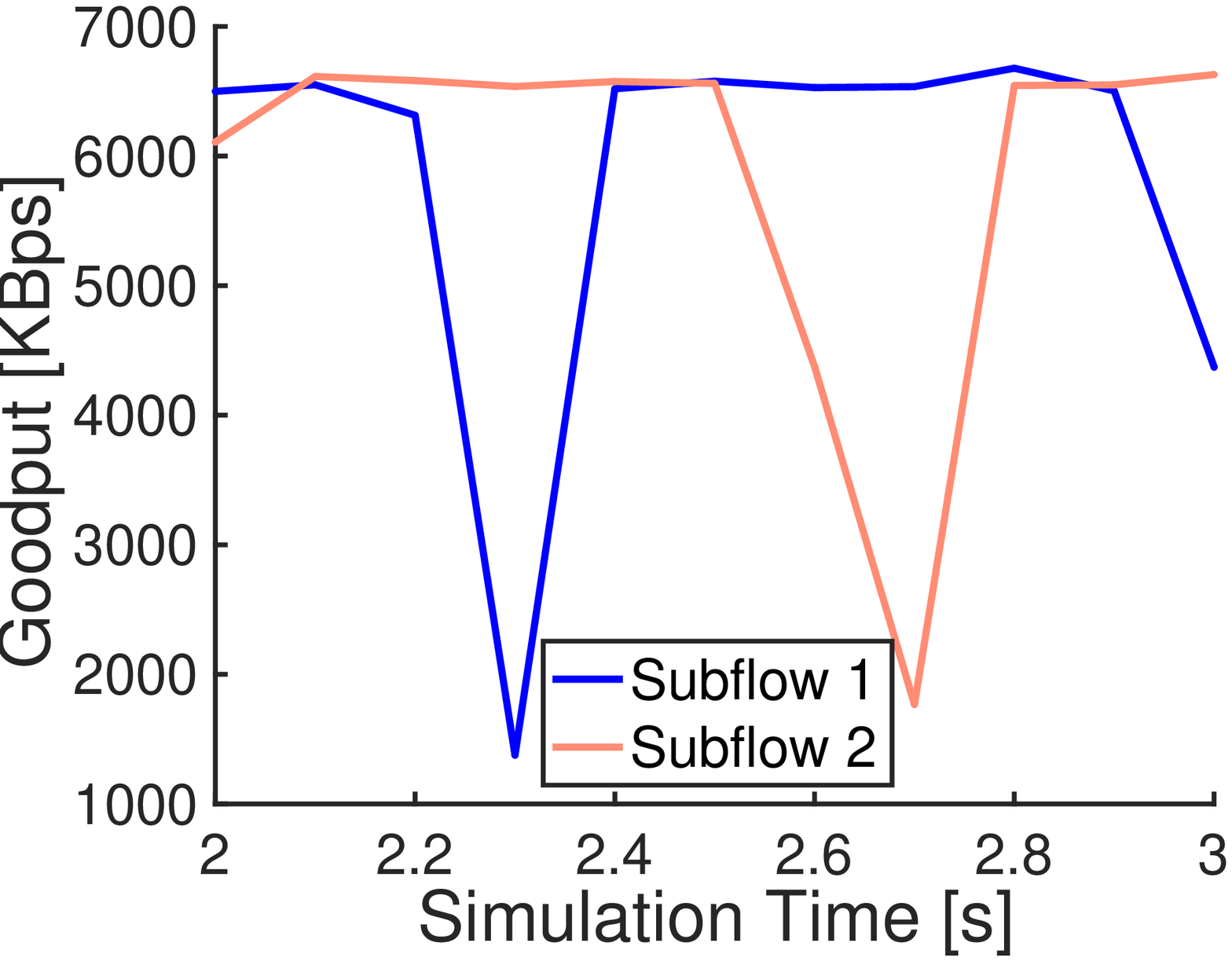}
  \caption{\label{fig:srttsched_throughput_compare}minSRTT}
\end{subfigure}
\caption{Comparison of goodputs of subflows obtained by QueueAware and minSRTT}
\label{fig:goodput_compare}
%\vskip -3mm
\end{figure}

\begin{figure}[!t]
\centering
\begin{subfigure}{.23\textwidth}
  \centering
  \includegraphics[width=\linewidth]{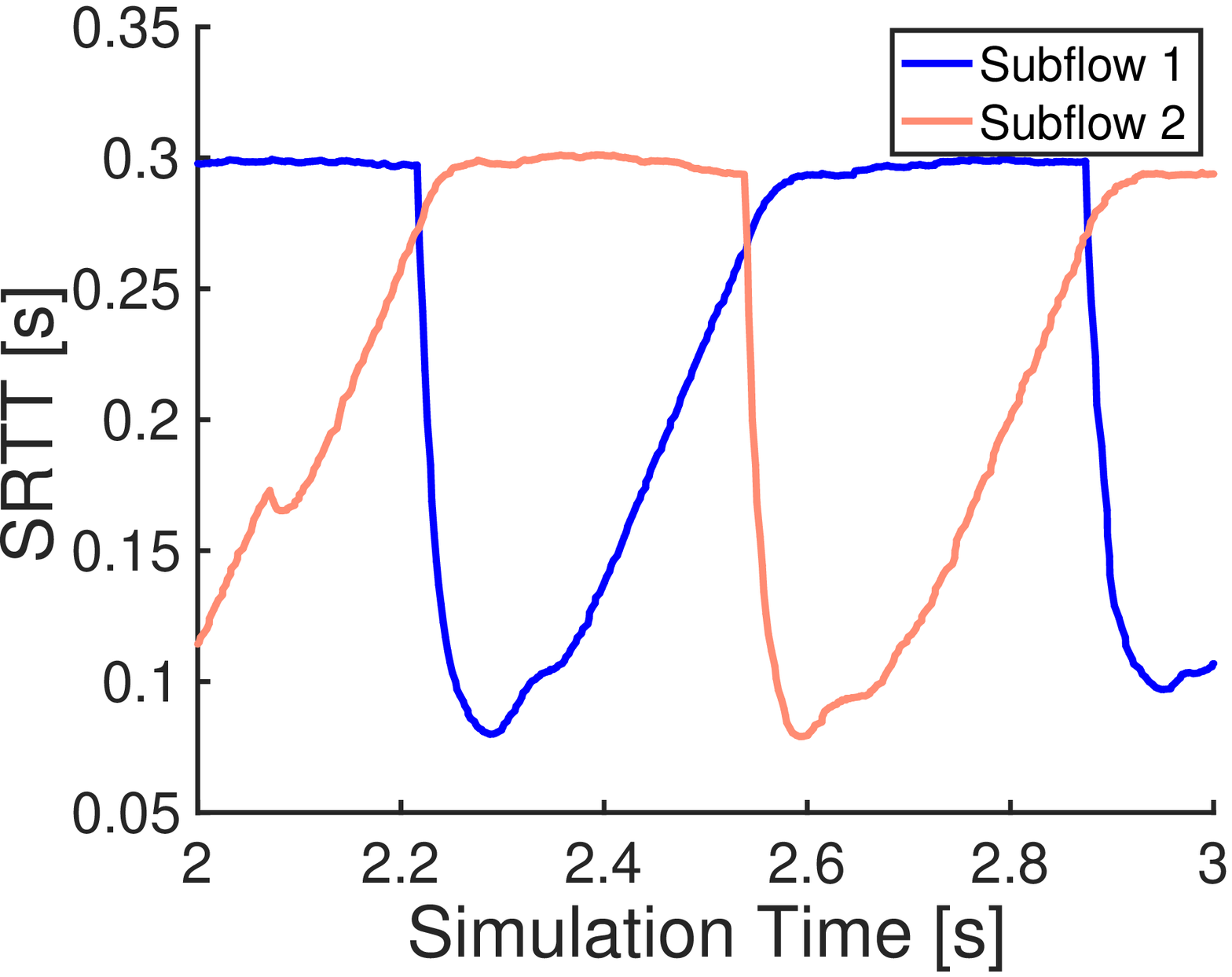}
  \caption{\label{fig:queuesched_srtt_compare}QueueAware}
\end{subfigure}%
\hspace{0.005\textwidth}
\begin{subfigure}{.22\textwidth}
\centering
  \includegraphics[width=\linewidth]{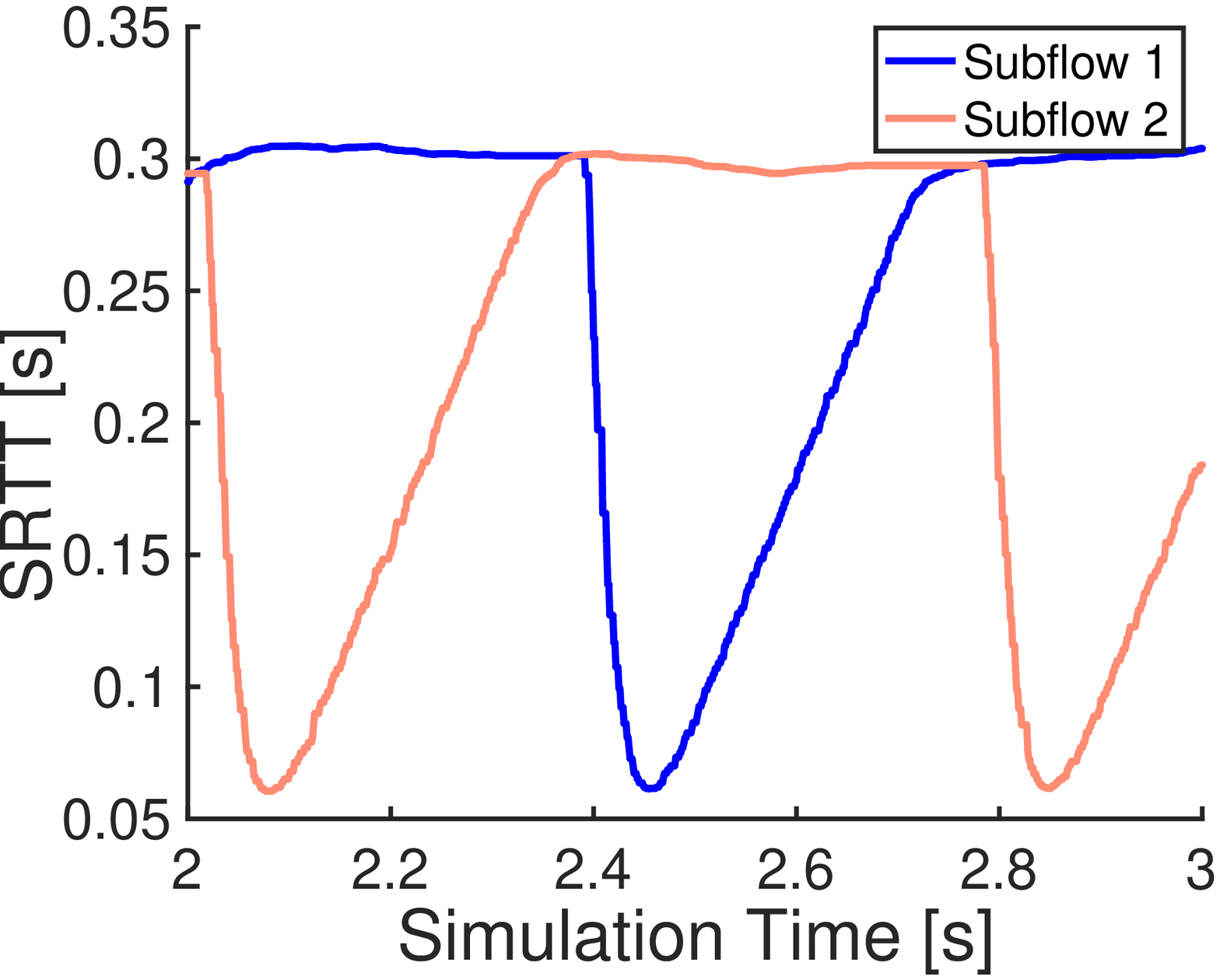}
  \caption{\label{fig:srttsched_srtt_compare}minSRTT}
\end{subfigure}
\caption{Comparison of SRTT of subflows obtained by QueueAware and minSRTT}
\label{fig:srtt_compare}
%\vskip -3mm
\end{figure}

\begin{figure}[!t]
\centering
\begin{subfigure}{.23\textwidth}
\centering
  \includegraphics[width=\linewidth]{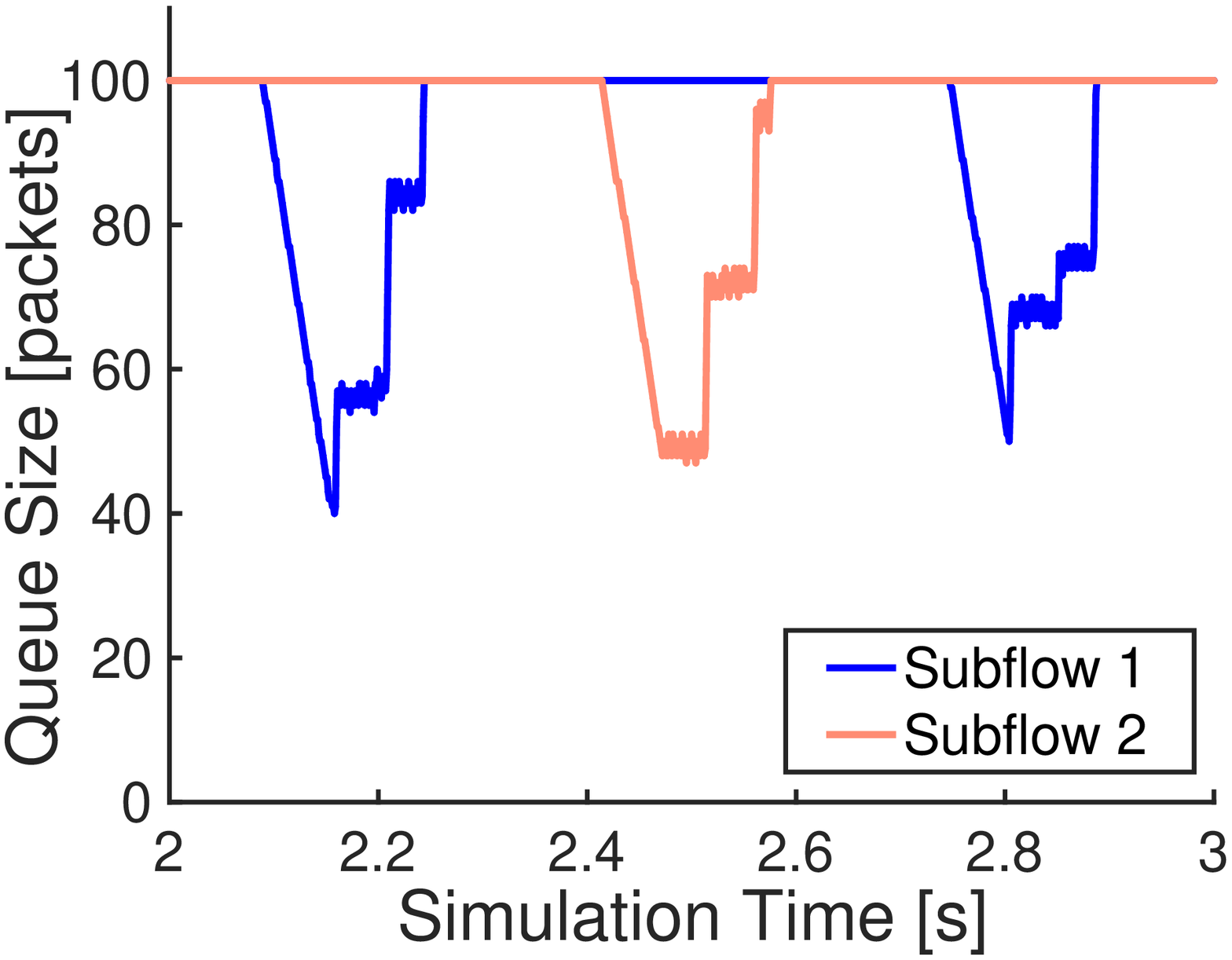}
  \caption{\label{fig:queuesched_queue_compare}QueueAware}
\end{subfigure}%
\hspace{0.005\textwidth}
\begin{subfigure}{.22\textwidth}
\centering
  \includegraphics[width=\linewidth]{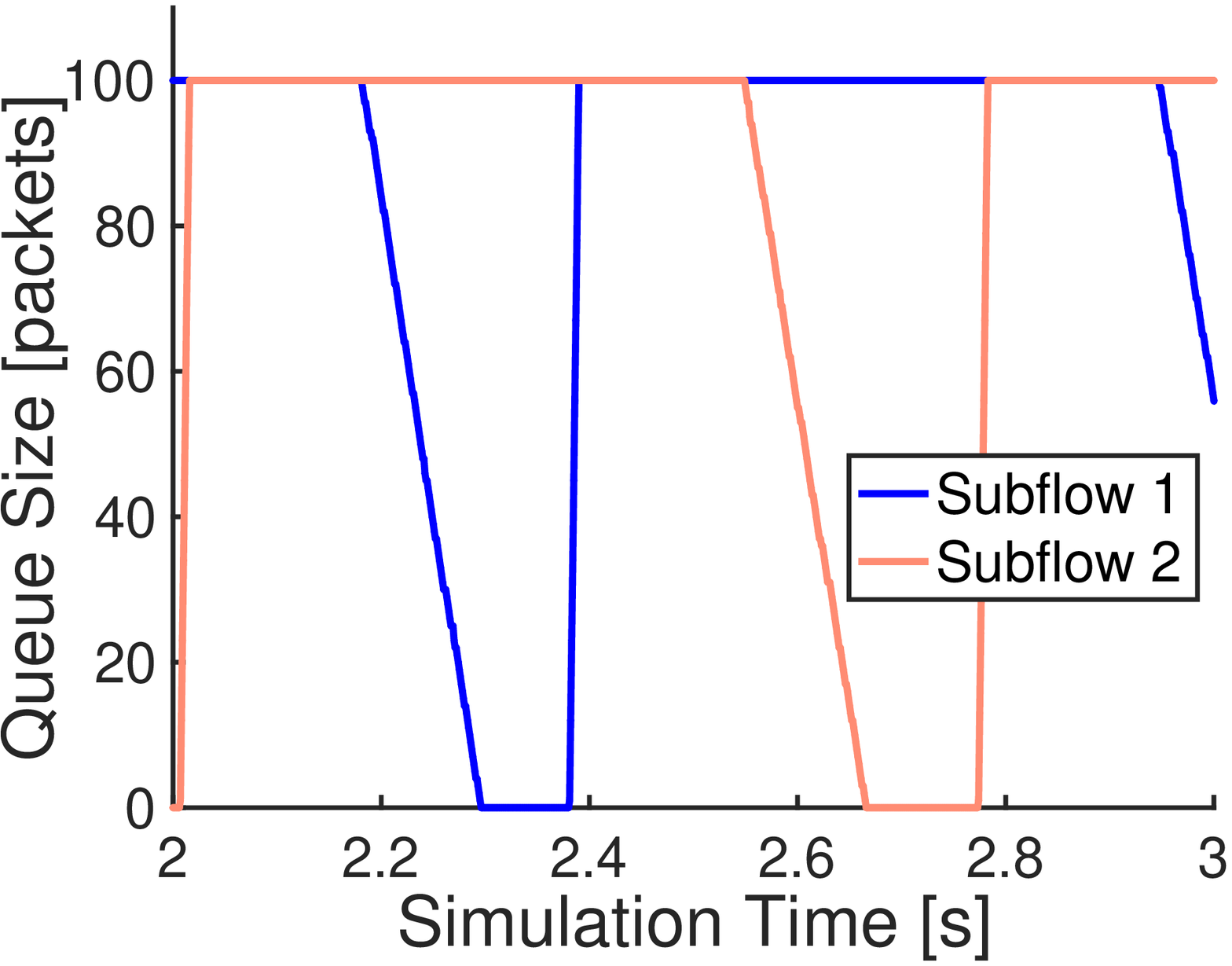}
  \caption{\label{fig:srttsched_queue_compare}minSRTT}
\end{subfigure}
\caption{Comparison of queue lengths of subflows obtained by QueueAware and minSRTT}
\label{fig:queue_compare}
%\vskip -3mm
\end{figure}

%Figure~\ref{fig:goodputs_result} shows the goodputs obtained  by the two schedulers on their two subflows for the different scenarios. 
We consider the scenario in which both TCP subflows use reliable paths. Figure~\ref{fig:goodput_wlan+wlan} compares subflow goodputs for when there are two identical WiFi access points. Observe that both QueueAware and minSRTT show similar behavior in total goodputs per flow. However, a larger goodput per flow is achieved by QueueAware. As a result, it achieves a $46\%$ increase in aggregate goodput over minSRTT. This is because minSRTT, as a consequence of only using the delayed feedback provided by SRTT, ends up scheduling packets to a subflow, which has a high device queue occupancy, for a longer time. Next, we shed more light on the differences in the behaviors of the schedulers, in Figures~\ref{fig:queuesched_throughput_compare}-~\ref{fig:srttsched_queue_compare}.

Figures~\ref{fig:queuesched_throughput_compare} and~\ref{fig:srttsched_throughput_compare} show the goodputs on the two flows obtained respectively by QueueAware and minSRTT as a function of time. While QueueAware can maintain a stable and almost equal goodput over both the flows, minSRTT at any given time chooses one flow over the other. This behavior of goodputs is made clear by Figures~\ref{fig:queuesched_srtt_compare} and~\ref{fig:srttsched_srtt_compare} that show SRTT behavior in same time duration. Subflows experience high SRTT for longer stretches of time when using the minSRTT scheduler, compared to when using the QueueAware scheduler. This corresponds to longer stretches of full queue occupancy of $100$ packets when using minSRTT (see Figure~\ref{fig:srttsched_queue_compare}), when compared to the QueueAware scheduler (see Figure~\ref{fig:queuesched_queue_compare}).

Figure~\ref{fig:goodput_wlan+wlan_6_12} shows the goodput performance when subflows use non-identical WiFi access points. It can be observed that QueueAware and minSRTT both perform similarly for the subflow that uses the faster link of $12$ Mbps. However, importantly, QueueAware makes good use of the other available $6$ Mbps link. Under QueueAware, the subflow using the slower link gets about $50\%$ of the goodput obtained by the subflow using the faster link. On the other hand, minSRTT gets hardly any goodput on the subflow using the slower link. The use of locally available cross-layer device queue occupancy information together with SRTT enables QueueAware to make good use of both the available WiFi interfaces.

Finally, Figure~\ref{fig:goodput_wlan+lte} shows the goodput performance when one of the subflows uses a WiFi interface and the other uses a $4$G interface that has twice the WiFi link rate. Recall that the $4$G network suffers from larger RTT. Even in this case we observe that QueueAware utilizes bandwidth offered by both network interfaces considerably better than minSRTT. QueueAware achieves $26\%$ more aggregate goodput when compared to minSRTT. The scheduler achieves a goodput that is $2.5\times$ that of minSRTT via $4$G. The underutilization of $4$G by minSRTT has earlier been observed in real-world deployment tests. It has been found that the default scheduler utilizes the WiFi sub-flow $71\%$ of time~\cite{realappsmptcp}.

%%--SRTT line plot--%%

%%--queue length line plot--%%
%The results show that QueueAware achieves \textasciitilde46\% increase in aggregated goodput of both flows compared to minSRTT. SRTT of a sub-flow provides delayed estimate of the network congestion utilising which minSRTT injects a large number of packets over already congested flow. The scheduler keeps sending packets on the sub-flow until delay due to congestion is reflected in SRTT and it surpasses that of the other flow. This leads to a step-wise flow usage where the sub-flows are active in a cyclic fashion. This effect can be observed in figure \ref{fig:srttsched_throughput_compare} and \ref{fig:srttsched_srtt_compare}.
%On the other hand, QueueAware scheduler considers the device queue utilisation along with SRTT to schedule packets on a sub-flow. As shown in figure \ref{fig:queueaware_compare}, the decision metric of QueueAware can detect flow congestion early and switches to sub-flow with lower congestion. It thus has more consistent sub-flow usage than SRTT and can achieve consistent goodput on both flows. 
\vspace{2mm}

\subsection{Unreliable Paths} 
On detecting packet loss, the congestion window of a sub-flow initiates the \textit{congestion avoidance} phase. This limits the number of packets that can be sent on the sub-flow. Furthermore, MPTCP employs \textit{Penalization and Retransmission (PR)}, where on packet error MPTCP reduces the congestion window of the sub-flow with high RTT and reinjects the lost packet on the other available subflow~\cite{mptcpietf}. Though this technique reduces the possibility of Head-of-Line (HoL) blocking, it also limits the sending rate and significantly impacts overall goodput. 

Figure~\ref{fig:error_2} shows the goodputs achieved by the two subflows when one of the subflows (subflow $1$ in figure) experiences a packet loss rate of about $10^{-2}$. Due to the above stated reason both schedulers achieve rather small goodputs on the subflow experiencing errors. However, QueueAware is able to better exploit the subflow that has a reliable path. On this path, it achieves a goodput that is an improvement of about $100\%$ over the goodput achieved by minSRTT.

%%%\begin{figure}[!t]
%%%\centering
%%%\begin{subfigure}{.28\textwidth}
%%%\centering
%%%  \includegraphics[scale=0.18]{Figs/through_error_2}
%%%  \caption{\label{fig:error_2}}
%%%\end{subfigure}
%%%\hspace{-0.1\textwidth}
%%%\begin{subfigure}{.28\textwidth}
%%%\centering
%%%  \includegraphics[scale=0.18]{Figs/through_error_both}
%%%  \caption{\label{fig:error_3}}
%%%\end{subfigure}%
%%%\caption{(a) error one (b) error both}
%%%\label{fig:error}
%%%%\vskip -3mm
%%%\end{figure}

\subsection{Small File Upload}
\begin{table}[t]
\centering
\scalebox{0.75}{
\begin{tabular}{@{}cccc@{}}
\toprule
                   &                             & \textbf{minSRTT}      & \textbf{QueueAware}      \\ \midrule
\textbf{Scenario}  & \textbf{Path Reliability} & \multicolumn{2}{c}{\textbf{Completion Time (s)}} \\ \midrule
Two WiFi, each rate $6$ Mbps                   & Reliable paths                & 1.456                 & \textbf{1.327}           \\
Two WiFi, each rate $6$ Mbps & Errors on one path   & 2.527                 & \textbf{2.204}           \\
                   %& Tx errors on both paths     & 2.695                 & \textbf{2.454}           \\ \midrule
WiFi and $4$G   & No Tx errors                & 2.439                 & \textbf{1.812}           \\ \bottomrule
\end{tabular}}
\caption{Small file upload completion time}
\label{tbl:app_compl}
\end{table}

Table \ref{tbl:app_compl} shows the upload completion time of a $10$ MB file for QueueAware and minSRTT for different interfaces and different path reliability. QueueAware achieves about $10\%$ decrease in upload time with respect to minSRTT.

%% file: relatedwork.tex
\section{Related Work}\label{sec:relatedwork}

Several researchers have proposed improvements to the default minSRTT scheduler of MPTCP. 
Paasch et al. \cite{paasch-scheduler} designed a modular scheduler framework for MPTCP and had compared the performance of default minSRTT scheduler with a Round-Robin scheduler. Baidya et al. \cite{slow-path-adaptation} propose RTT-based path quality metric to adapt out-of-order transmissions while limiting the usage of a slower path.
Kuhn et al. \cite{daps} aim to reduce overall application delay by estimating maximum allowed receiver buffer blocking time to transmit out-of-order packets on multiple paths.
On the other hand, Hwang et al. \cite{packet-scheduling} propose to freeze the utilization of slower path when the difference in RTT's of faster and slower paths exceeds a calculated threshold. BLEST \cite{blest} and OTIAS \cite{otias} balance heterogeneous flows and reduce Head-of-Line blocking by considering several parameters such as CWND, in-flight packets etc., along with SRTT.
CMT-RMDS \cite{receiver-scheduler} proposes adopting receiver-centric path characteristics along with sender-driven RTT values to better estimate current path conditions. 

Researchers have also proposed to utilize other network parameters to provide better path estimation. 
Corbillon et al. \cite{crosslayer_video} leverage application layer information in transport layer flow scheduling decisions to provide delay-resilient video streaming in MPTCP. 
Lim et al. \cite{crosslayer_infocomm} labels WiFi subflow as active/inactive for data transmission based on a minimum desired signal strength.
F2P-DPS \cite{f2dpds} proposes to combine several TCP parameters such as CWND, SSThresh, RTTs to estimate subflow weights for data data transmissions. 
Ni et al. \cite{ocps} utilizes reverse-path SACK packets to inform the sender of any out-of-order/lost packets at receiver buffer and calculate offset to provide successful data chunk delivery.

Although the solutions mentioned above tackle several critical issues affecting Multipath TCP in real-world, these techniques are significantly dependent on accurate estimation of current path characteristics. The solutions which utilize SRTT value as current path performance metric or are dependent upon it suffer from same issues as that of minSRTT shown in the paper. To efficiently handle varying application data traffic over heterogeneous paths in MPTCP, an ideal scheduler must be able to schedule packets over flows proactively and must adapt to network conditions swiftly. 
%Furthermore, some of the works either propose to change the inherent architecture of MPTCP protocol or introduce customized packets in-order to facilitate their solution which we think . 

%% file: discussion.tex
\section{Discussion and Future Work}\label{sec:discussion}

\begin{table}[!tb]
\centering
\scalebox{0.75}{
\begin{tabular}{@{}ccc@{}}
\toprule
\textbf{Host Machine}  & \textbf{Network}        & \textbf{Receiver}      \\ \midrule
Data rate  & Channel utilization     & Bandwidth utilization  \\
Retry percentage       & Path congestion         & Receiver queue delays  \\
Network interface type & Number of nodes on path & Congestion window size \\ \bottomrule
\end{tabular}
}
\caption{Network parameters impacting RTT}
\label{network-parameters}
\end{table}

The aim of our evaluation was to show that RTT alone is not the best
measure for scheduling packets over multiple network
interfaces in MPTCP.
 Sender device queue is one such parameter which needs to be monitored and considered in overall network scheduling decision.
Only by including cross-layer feedback into MPTCP control loop can we fully exploit the potential offered by multiple connections.
As shown in the paper, treating the links as independent, parallel TCP flows restrict MPTCP's performance.
%  This urges MPTCP researchers to contemplate cross-layer feedback
% loops and external network parameter while scheduling data packets
% over TCP subflows.

Several other conditions can lead to changes MPTCP's efficiency. Table~\ref{network-parameters} lists few such variables that impact RTT and the network path and which could be exploited by a more holistic MPTCP. These include host-specific parameters, such as device queue length; network-specific aspects, such as congestion, and receiver-specific parameters, such as window size. An efficient MPTCP should look at all or many of these parameters in conjunction while deciding how to schedule the packets and allocate the available flows for data transport.

Several other factors can also significantly impact network behavior. Congestion at an access point, over-utilization of channel capacity, queueing delay at receiver are some such scenarios. However, unlike the locally accessible parameters (such as device queue occupancy), explicitly monitoring and predicting these parameters is an interesting research question of its own. One possibility is to use explicit notifications such as ECN to convey occurrence of such a scenario. Incorporating these mechanisms and comparing its performance impact on current MPTCP implementation would be an interesting future avenue for research in the wider community.

In this paper, we argue that the current way of treating the individual subflows, practically separately relying only on their TCP level information, is not sufficient for fully exploiting the potential offered by multiple connections, and a more holistic approach to MPTCP and how it manages the connections is needed. Our results in this paper illustrate only one small facet of the bigger problem, but broader research efforts are needed to understand better how such multiple connections should be used in modern networks. These cover questions such as fairness to other network flows, general performance issues, flow control, and security issues, but they all apparently require a different approach to the problem than parallel, but largely independent connections.

%% file: conclusion.tex
\section{Conclusion}\label{sec:conclusion}

In this paper, we demonstrated the shortcomings of MPTCP scheduling algorithm and the inadequacy of its reliance on SRTT. We proposed the QueueAware scheduler, which exploits device driver queue occupancy together with sRTT to obtain significantly better performance than the default MPTCP scheduler. We believe that the QueueAware scheduler highlights the need for a more holistic approach to multipath scheduling than currently being done by MPTCP.% Research is needed for a better understanding how multipath connections should be managed and used, to get better performance and better utilization of the network resources.

%In this paper, we showed the gap in subflow management by Multipath TCP. The current implementation reliance on TCP-level information such as SRTT and congestion window leads to sub-optimal usage of available flow potential. We thus presented QueueAware, a scheduler for MPTCP which considered device layer queue occupancy for injecting TCP packets in subflows. We implemented QueueAware scheduler model in NS3 and compared it with the default MPTCP scheduler, minSRTT. We compared the performance of both schedulers in several realistic WiFi and cellular use cases. We observed that QueueAware can achieve more that 45\% increase in application goodput by incorporating cross-layer information. Finally, we provided a set of cross-flow parameters and network conditions that MPTCP must consider in order to have a more holistic view of it associated network. 